# Are cryptocurrencies real financial bubbles?

# Evidence from quantitative analyses

**Marco Bianchetti**, Intesa Sanpaolo, Financial and Market Risk Management, and University of Bologna[1]
**Camilla Ricci,** Intesa Sanpaolo, Financial and Market Risk Management
**Marco Scaringi**, Intesa Sanpaolo, Financial and Market Risk Management

**Abstract**

The growth of peer-to-peer exchanges and the blockchain technology has led to a proliferation of cryptocurrencies and to a massive increase in the number of investors who actually negotiate digital money. Cryptocurrencies trade at prices mainly driven by investor sentiment, becoming a potential source of financial bubbles and instabilities.

In this work, we apply quantitative models to the study of Bitcoin and Ether, two of the most famous cryptocurrencies. Our bubble detection methodology combines the Log Periodic Power Law (LPPL) model, originally created by Johansen, Ledoit and Sornette (JLS), and the statistical model developed by Phillips, Shi, and Yu (PSY). In particular, we employ three different versions of JLS model, i.e. Ordinary Least Square (OLS), Generalised Least Squares (GLS) and Maximum Likelihood Estimation (MLE), and two PSY statistical tests (BSADF and BSADF*).

We find that, during the sample period 1st December 2016 – 16th January 2018, Bitcoin shows typical hallmarks of a bubble phase in mid December 2017 and in the first half of January 2018, anticipating the large crashes observed thereafter. Also the Ether price dynamics reveals bubble evidence in mid June 2017, anticipating the crash observed on 12th June, and a weaker signal around 12th January 2018, anticipating the crash observed in the same days.

This paper confirms the high risk of speculative bubbles associated with cryptocurrencies, related to investor exuberance pumping market prices far away from their fundamental values, thus creating critical situations subject to possible crashes. Our methodology is general and can be applied to virtually any financial time series, and may support investing and risk management strategies.

---

[1] Corresponding author, marco.bianchetti(at)unibo.it.

**Table of contents**

**Acknowledgments**
The authors gratefully acknowledge Massimo Morini and other colleagues in Financial Market Risk Management of Intesa Sanpaolo for fruitful discussions and insights into digital currencies, and Luca Lopez, Angelo Salvatori, Francesco Reggiani and Davide Emilio Galli for common efforts at previous stages of our work on bubble analysis.

**Disclaimer**
The views and the opinions expressed in this document are those of the authors and do not represent the opinions of their employers. They are not responsible for any use that may be made of these contents. The opinions, forecasts or estimates included in this document strictly refer to the document date, and there is no guarantee that future results or events will be consistent with the present observations and considerations. This document is written for informative purposes only, it is not intended to influence any investment decisions or promote any product or service.

## 1. Introduction

Since Bitcoin brought cryptocurrencies into the spotlight in 2009, the number and diversity of digital coins expanded dramatically [1], [2]. There are two key innovations behind cryptocurrencies: the technology employed and the way they are governed. The first innovation, technology, is the blockchain, a public ledger that contains all transaction records since the inception of the currency, with the name of users stored under their public addresses (which are strings of characters akin to bank account numbers). The second innovation of cryptocurrencies is its decentralized governance [3].

The extraordinary returns generated by cryptocurrencies such as Bitcoin has led to a frenzy of investment activity and interest from traditional investors. However, most of the established and highly traded cryptocurrencies, such as Bitcoin and Ether, qualified as commodities rather than securities, and thus not subject to securities laws. Hedge funds that trade in these cryptocurrency commodities, or "crypto funds," fall almost entirely outside the extensive securities regulations that would apply to traditional hedge funds [4].

Cryptocurrencies, unlike other alternative investment classes that have real assets, are digital assets and their fundamental value is hard to comprehend. Their store of value can be related to the development of the technology underneath, such as blockchain, and to the computing power engaged in mining. As a result, the cryptocurrency market is mainly driven by investor sentiment, leading to a high volatility [5]. Holders of large amounts of Bitcoin are often known as whales. They can send prices plummeting by selling even a portion of their holdings. About 40 percent of Bitcoin is held by perhaps 1,000 users, they can coordinate their moves or preview them to a select few. Bittrex, a digital currency exchange, recently wrote to its users warning that their accounts could be suspended if they banded together into "pump groups" aimed at manipulating prices [6].

There's no agreed authority for the price of Bitcoin, and quotes can vary significantly across exchanges. For example, In Venezuela, where inflation is skyrocketing and basic necessities run in short supply, many took to Bitcoin mining despite government efforts to crack down on miners [7]. In Zimbabwe, following a lack of confidence in the local financial system, the cryptocurrency has traded at a persistent premium over $10,000. Russia is preparing to issue a government-backed cryptocurrency, the CryptoRuble. Unlike decentralised cryptocurrencies such as Bitcoin and Ether, there is no mining involved in CryptoRuble – all transactions are recorded via blockchain and verified by a centralised government authority [8].

Prices and volumes are difficult to assess. Bloomberg publishes a price that draws on several large Bitcoin trading venues [9]. The CME CF Bitcoin Reference Rate (BRR) provides a once-a-day reference rate for the US dollar price of Bitcoin which aggregates the trade flow of major Bitcoin spot exchanges during a calculation window into the US dollar price of one Bitcoin. The BRR is designed around the Iosco principles for financial benchmarks. Bitstamp, GDAX, itBit and Kraken are among the exchanges currently contributing the pricing data for calculating the BRR [10].

Lately, the sharp rise in Bitcoin has sparked analogies to past irrational financial exuberance, from the tulip mania to the dotcom boom and bust. In fact, in the middle of December, Bitcoin surpassed $20,000 in a matter of hours, taking 2017's price surge to almost 12-fold as buyers shrugged off increased warnings that the largest digital currency is an asset bubble. A possible explanation for the appeal of Bitcoin is also given by the limited nature of supply, as new coins can only be created through complex calculations, and the total number of coins is limited to 21 million. Even as analysts disagree on whether the largest cryptocurrency by market capitalization is truly an asset, its $170 billion value already exceeds that of about 95 percent of the S&P 500 Index members [11].

A recent study investigated the behaviour of cryptocurrencies time series, finding that their dynamics is quite complex, displaying extreme observations, asymmetries and several nonlinear characteristics which are difficult to model. Differently from foreign exchange currencies, leverage effect has a substantial contribution in the volatility dynamics. On average, volatility increases more after negative shocks than after positive shocks as in the equity market, hence, crypto–currencies time–series incorporate the so–called leverage effect with some degree of heterogeneity across the series [12].

Until December 2017 no derivatives were available on Bitcoin, when CBOE and CME Group launched Bitcoin futures [13]. Bitcoin exchange-traded funds seem to agree on the fact that once a derivative is launched, a Bitcoin ETF will follow [14].

While most global banks support the use of distributed ledgers and blockchain with innovative projects, some bankers have explicitly opposed the idea of investing in Bitcoin. Most notably, JP Morgan chief executive Jamie Dimon called Bitcoin a fraud, and Nobel Prize-winning economist Joseph Stiglitz, along with a host of economists and financiers, denounced the crypto rally as a craze [15], [16]. However, according to other analysts, the introduction of financial instruments indexed to Bitcoin has the potential to elevate cryptocurrencies to an emerging asset class, such as gold. Bitcoin has not yet overtaken gold in terms of investment levels, total value in circulation or transaction volumes, but it has grown rapidly and could make more progress to match or even surpass the use of gold.

This paper is structured as follows: Section 2 describes the our methodology for financial bubble detection. Section 3 shows the procedures and analyses the results of our test for bubbles in the Bitcoin and Ether markets. Finally, Section 4 presents our conclusions.

## 2. A Look at Bitcoin and Ether

In Figure 1 we show some market data related to the two cryptocurrencies, i.e. prices, market capitalizations, exchange traded volumes, historical volatilities and correlations. Crypto-trading volumes are dominated by Bitcoin, but other cryptocurrencies, Ether first, are gaining more and more room. The Mt. Gox exchange prevailed until its bankruptcy in February 2014. Nowadays the market is fairly less concentrated and more liquid. The historical volatility in Figure 1 is calculated as standard deviation of logarithm returns, based on thirty daily closing prices (that is thirty business days for Bitcoin, gold spot price, EUR/USD exchange rate and thirty calendar days for Ether). Historical volatilities of Bitcoin and Ether are much larger (~10 times) than fiat currencies or gold. A factor that could help limit cryptocurrency volatility is an emergence of derivatives such that investors can hedge or short sell their positions by using Futures contracts [19]. Historical correlations between Bitcoin and Ether, fiat currency and gold are very variable along time, ranging from ca. $\pm$80%. Thus, adding digital currencies to a portfolio can act as a powerful source of diversification.

Regarding trading volumes, prior to the fall in February 2014, Mt. Gox, a popular Bitcoin exchange, dominated an estimated 80-90% of the Bitcoin-Dollar trades. The bankruptcy of Mt. Gox resulted in an alleged loss of 850,000 Bitcoin leaving the troubled exchange insolvent and many customers out of pocket [17]. The exchanges today claim to have learned from Mt. Gox and present themselves as advanced models with better security mechanisms in place. The trading volumes rise again which implies that the market is fairly liquid. The trading volume during the day is closely linked to the trading times the US and European stock markets [18] The biggest change, however, could continue to be the sharp decrease in Bitcoin’s dominance should Ether and other tokens capture a larger percentage of overall trading.

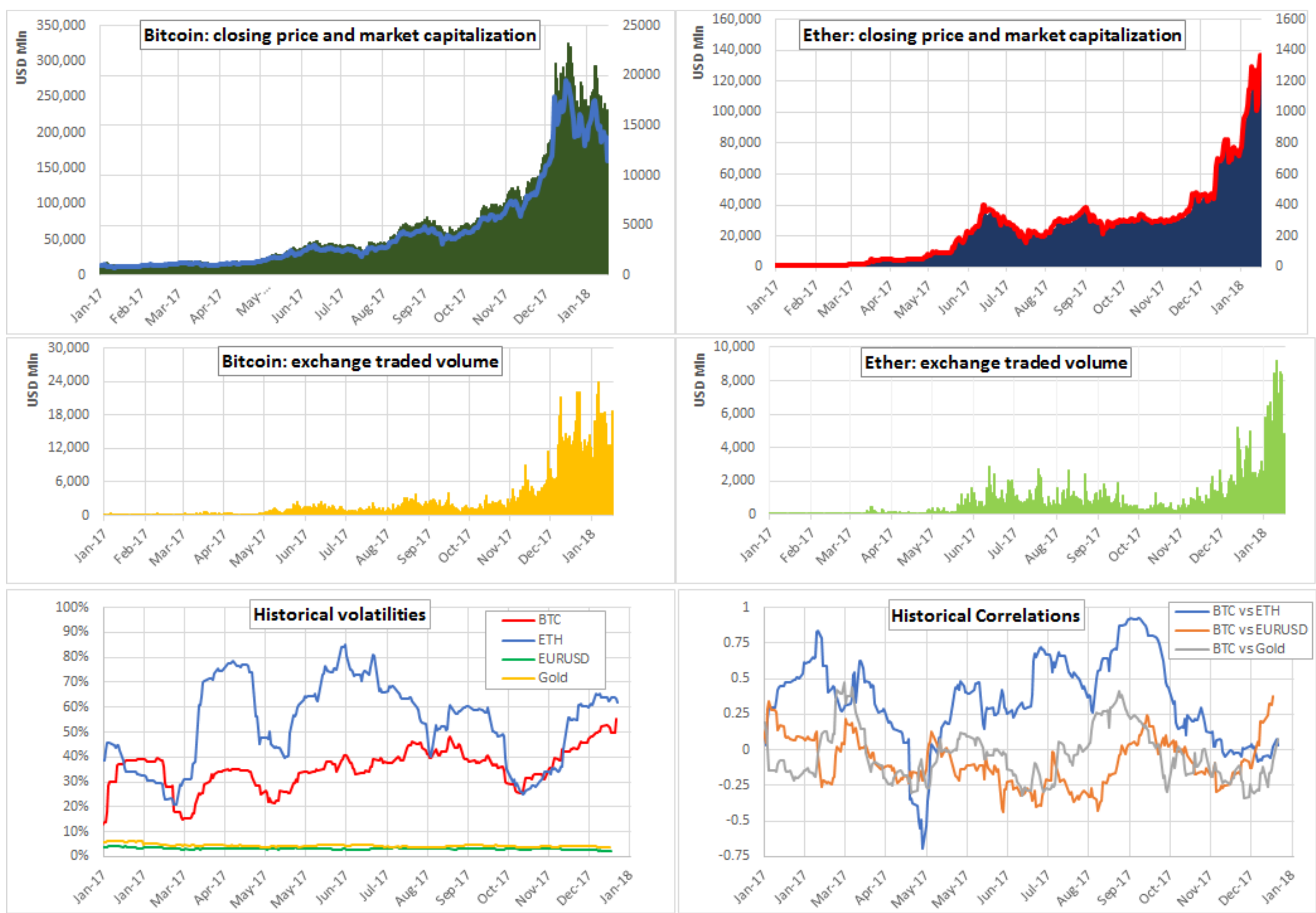

**Figure 1**. Top panels: prices and market capitalizations for Bitcoin (left) and Ether (right). Middle panels: exchange traded volumes. Source: Coinmarketcap. Bottom left panel: historical volatilities of gold spot price (from Bloomberg XAU BGN Curncy), EUR/USD exchange rate (from Bloomberg: EURUSD BGN Curncy), Bitcoin (from Bloomberg: XBTUSD Curncy) and Ether (from Etherscan), calculated with rolling windows including 30 market data, and normalized to 1 year. Bottom right panel: correlations between Bitcoin and Ether, EURUSD and Gold, calculated with rolling time windows including 30 daily returns.

## 3. Methodology

### 3.1. Quantitative Analysis of Financial Bubbles

Financial markets exhibit a complex organization and dynamics with individual market agents characterized also by herding behaviour, such as imitation, self-organized cooperativity and positive feedback. Such phenomena influence the dynamics of asset prices and may lead to the development of endogenous instabilities.

The origin of a positive bubble lies in the unsustainable pace of the asset market price growth fed by self-reinforcing over-optimistic anticipation. In a financial bubble, it is the expectation of future earnings rather than present economic reality that motivates the average investor and causes the divergence between the asset market price with respect to its fundamental value. The same dynamics appears in case of negative bubbles, where self-reinforcing pessimistic anticipation of future losses forces the asset market price below its fundamental value. During a positive (negative) bubble regime, the asset market price becomes unstable and very sensitive to exogenous factors, such as upcoming news with a negative (positive) consequence on the asset, that may lead to a collapse (rebound) of the asset market price.

The mainstream publications about modelling financial bubbles review econometric approaches, such as the model developed by Phillips, Shi, and Yu (PSY, see sec. 3.3). This model looks for a possible explosive behaviour of the time series with respect to its fundamental value using statistical tests, and is able to give indications about the initial and final dates of a period of exuberance.
Another stream of publications adopts a different approach relying on agent-based models, such as the model originally developed by Johansen, Ledoit and Sornette (JLS, see sec. 3.3). In this context, bubble regimes and large stock market crashes (rebounds) can be seen as the so-called critical phenomena studied in statistical physics in relation to magnetism, melting and other phase transitions. According to the JLS model a bubble regime is characterized by a super-exponential path generated by imitation and herding behaviour of investors and traders creating positive feedback in the market asset price that this critical point is decorated by oscillations describing instabilities generated by tension and competition between fundamentalists and noise traders. Such dynamics leads to a finite-time singularity at some future critical time, interpreted as the forecast of a possible crash or rebound for the asset price. As a proof, the implementation of this modelling technique is reported in several academic studies and is regularly applied by the Financial Crisis Observatory at ETH Zurich [29].

### 3.2. Previous Applications to Cryptocurrencies

Even if the industry of cryptocurrencies has grown exponentially over the past several months, to the best of our knowledge there are few applications of financial models to the study of this new markets.

MacDonell [20] used the JLS model to forecast successfully the Bitcoin price crash that took place on December 4, 2013, showing how the model can be a valuable tool for detecting bubble behaviour in digital currencies. Malhotra *et al*. [21] investigated the reasons behind the steep rise and fall of Bitcoin exchange rates in 2013-2014 applying the sequential ADF tests of PSY. The results from sequential root tests provided strong evidence of explosiveness in Bitcoin exchange rates even if the unreliability of data obtained from internet limits the accuracy of empirical evidence. Corbet *et al*. [22] found evidence of price bubbles in Bitcoin and Ether using the PSY model. Fantazzini *et al.* [23] reviews the econometric and mathematical tools which have been proposed to model the Bitcoin price and several related issues, highlighting advantages and limits.

### 3.3. Our Approach

In our study, we analyse two popular cryptocurrencies, Bitcoin and Ether, looking for clues of speculative bubbles in the time window that spans more than one year (from 1st December 2016 to 16th January 2018). The methodology used in this work is based on the two main approaches described in sec. 3.1 above: the statistical model originally created by PSY [24], [25], [26], and the Log Periodic Power Law (LPPL) model, originally created by Johansen, Ledoit and Sornette (JLS or LPPL model) [27], [28].

In order to test and qualitatively assess the confidence level of our findings, we employ different versions of JLS model, i.e. Ordinary Least Square (OLS), Generalised Least Squares (GLS) and Maximum Likelihood Estimation (MLE [29]). The JLS-GLS approach has been used in this work, to our knowledge, for the first time [32]. Also, the PSY model statistical tests come in two flavours, i.e. Backward Supremum Augmented Dickey Fuller (BSADF and BSADF*).

The calibration of the models to the market data is a critical numerical problem. For JLS we adopt a multiple time windows calibration strategy where, for each date in the time series, we run many calibrations with a variable window length. The procedure works as follows: for each historical series, we fixed a window length (72 points for Bitcoin, 69 points for Ether), which we moved forward with a one-point step. For each step, we test 20 windows keeping fixed the right extremum and changing the left extremum (thus ranging from 52 to 72 points for Bitcoin and from 49 to 69 points for Ether). Finally, for each single window we perform a calibration and count the number of valid bubble signals. The latter corresponds to JLS fits satisfying the so-called “Sornette bounds”, a set of constraints on JLS parameters established through both mathematical considerations and backtesting on known past bubbles [29].

Since the LPPL function is non-linear and oscillating, the corresponding fit function presents multiple local minima, where a naïve (local) optimization method would get trapped. Hence, we adopt a robust global multi-dimensional optimization approach based on Genetic Algorithms (GA). Such approach attacks the problem without any assumption on the shape of the fit function and leads to the global minimum corresponding to the optimal JLS parameters.

The PSY model requires, instead, an ADF test to calculate the statistics and a Monte Carlo simulation to calculate the relative confidence levels. The multiple time windows strategy is the same presented in [24], [32].

For both JLS and PSY implementations, we could reproduce previous results available in the literature [31], [32]. Since such calculations must be repeated for each combination of time series, calibration date and window, the resulting numerical effort is relevant and requires huge computational resources.

## 4. Results

In this section, we present the results of our analysis with JLS and PSY models applied to Bitcoin and Ether, aiming at determining whether they can represent a source of financial instabilities.

### 4.1. Bitcoin

The daily Bitcoin data is collected from Bloomberg (XBTUSD Curncy), where the price of one BTC in USD is given by an average from the world's leading Bitcoin exchanges. The time window analysed spans from 1st December 2016 to 16th January 2018.

The results obtained for Bitcoin are shown in Figure 2. We observe that both JLS and PSY models coherently find strong bubble signals. In particular, PSY shows that the bubble period started in May 2017 and is still alive in January 2018. JLS predicts several critical times consistently across the three versions (OLS, GLS, MLE). A first clear group of signals appears around the half of December 2017, anticipating the large crash (~30%) observed in the second half of the month. A second broader group appears thereafter, anticipating the other flash crash observed in the first half of January and possibly other crashes in the future (we remind that our data set stops at 16th January 2018). These crashes can be related to recurring negative news of China and South Korea regarding restrictions for cryptos. We also notice that JLS does not reveal valid bubble signals around the slump observed in early September 2017 (~28% in two weeks), hence we may look at it as a local swing related to exogenous information flow (the Chinese restrictions on ICOs).

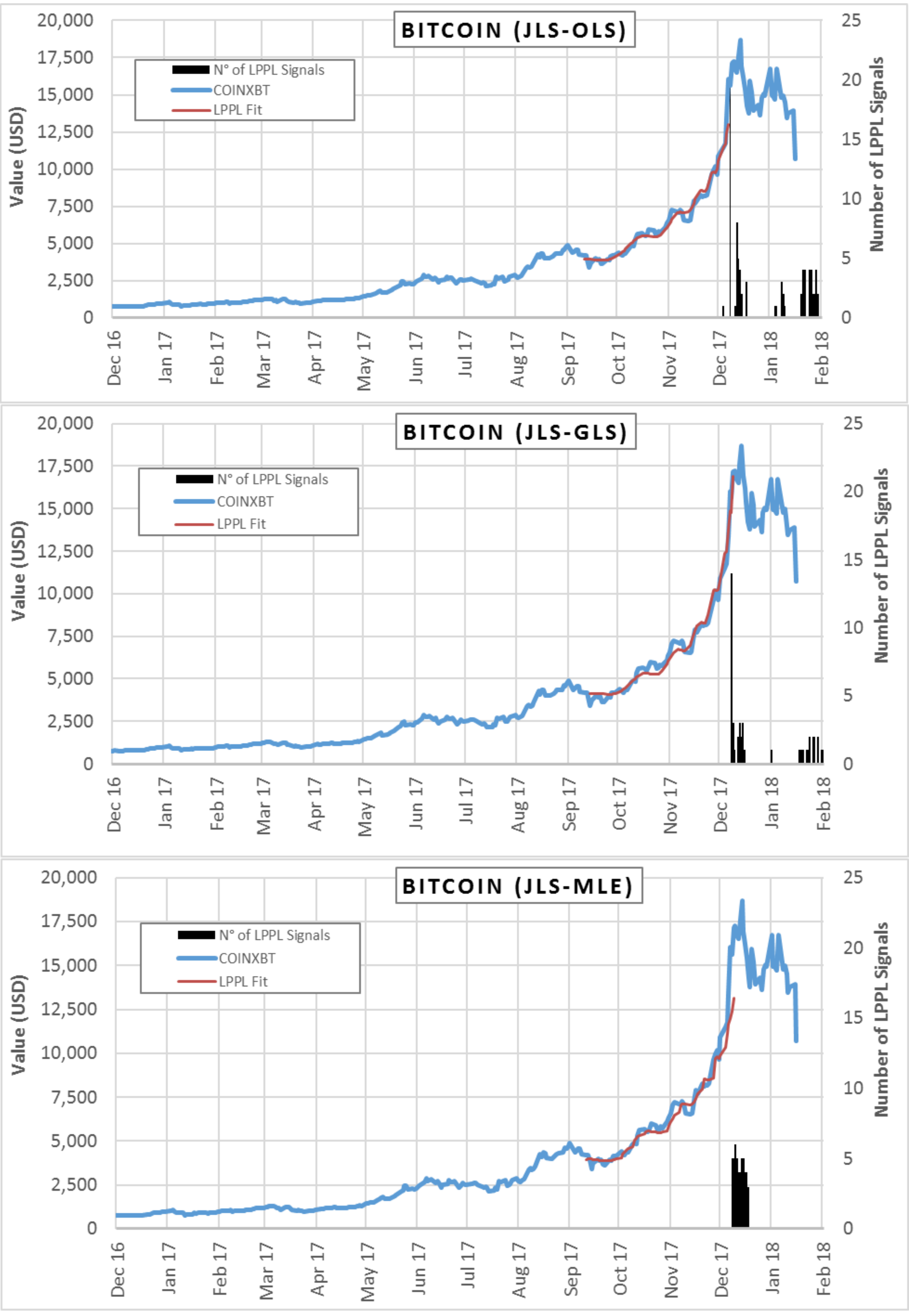
BITCOIN (JLS-OLS)
N° of LPPL Signals
COINXBT
LPPL Fit
Value (USD)
Number of LPPL Signals
20,000
17,500
15,000
12,500
10,000
7,500
5,000
2,500
0
25
20
15
10
5
0
Dec 16
Jan 17
Feb 17
Mar 17
Apr 17
May 17
Jun 17
Jul 17
Aug 17
Sep 17
Oct 17
Nov 17
Dec 17
Jan 18
Feb 18
BITCOIN (JLS-GLS)
N° of LPPL Signals
COINXBT
LPPL Fit
Value (USD)
Number of LPPL Signals
BITCOIN (JLS-MLE)
N° of LPPL Signals
COINXBT
LPPL Fit
Value (USD)
Number of LPPL Signals

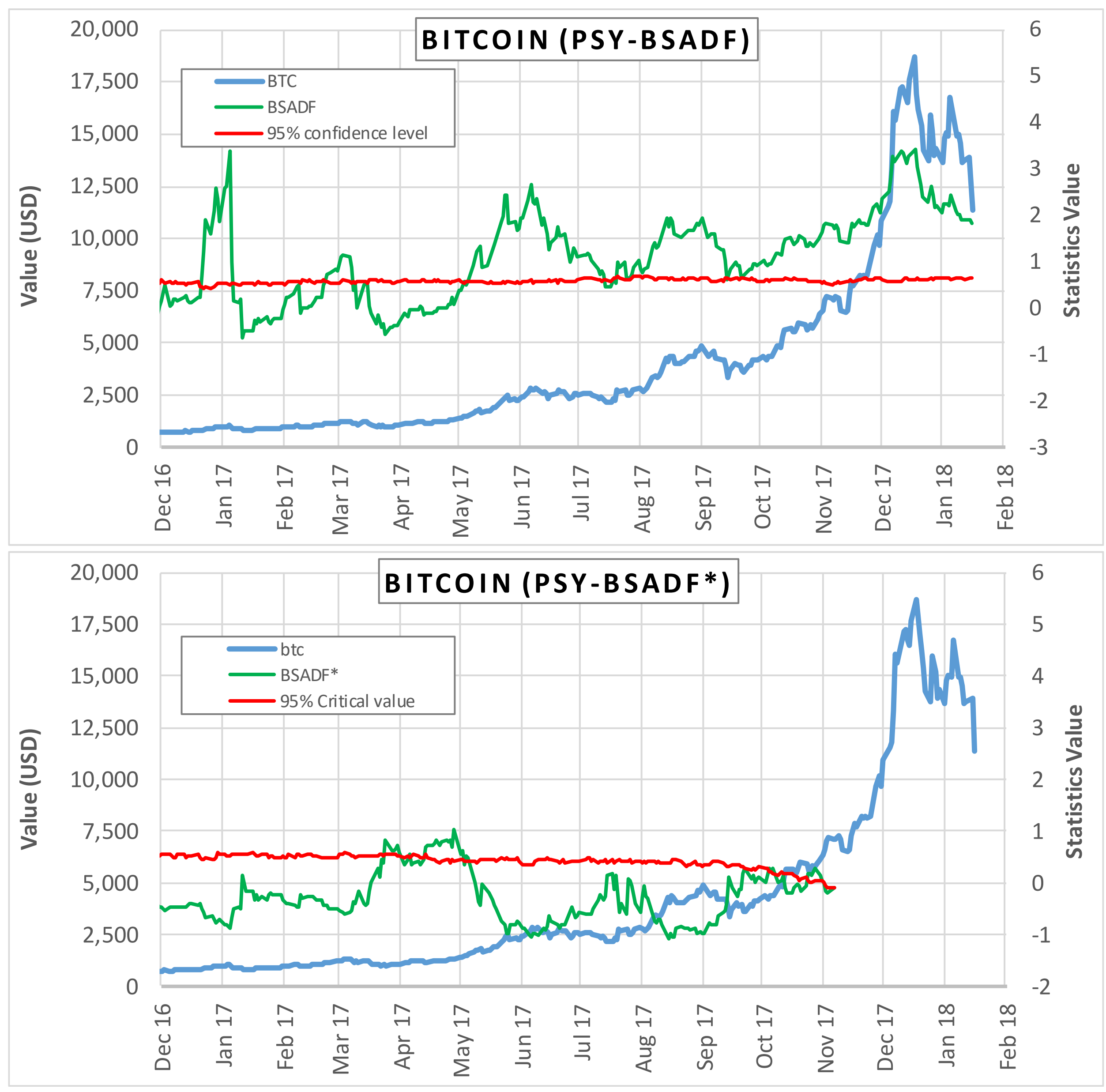

***Figure 2**: Bitcoin results: the first three panels starting from the top show JLS model results for each calibration technique (OLS, GLS and MLE). The left axis refers to the asset value (blue), while the right axis counts the number of LPPL signals represented by bars (black) in correspondence of the critical time $t_c$. One LPPL fit is also shown (red) as an example of valid JLS signal. The last two panels at the bottom show the PSY model results. The left axis refers to the asset value (blue), while the right axis refers to the BSADF and BSADF* statistics (green, fourth and fifth panels) and to the 95% critical value of the statistics (red). See Table 1 for chart interpretation. Source: Bloomberg XBTUSD Curncy.*

### 4.2. Ether

The Ether time series (ETH) is downloaded from Etherscan, an exchange supplying daily prices, in the time window from 1st December 2016 to 16th January 2018. Ether is a token provided by Ethereum platform, an open source blockchain-based distributed computing platform which was formed in 2016 after the fork between "Ethereum Classic" (ETC) and "Ethereum" (ETH) [33].

In Figure 3 we show in the results obtained for Ether. Also in this case, both JLS and PSY coherently find bubble signals. In particular, PSY shows two distinct bubble periods, one in March-September 2017, and another one started in December 2017 and still alive. All the three versions of JLS coherently find a bubble signal around mid-June 2017, corresponding to the crash observed on 12th June.

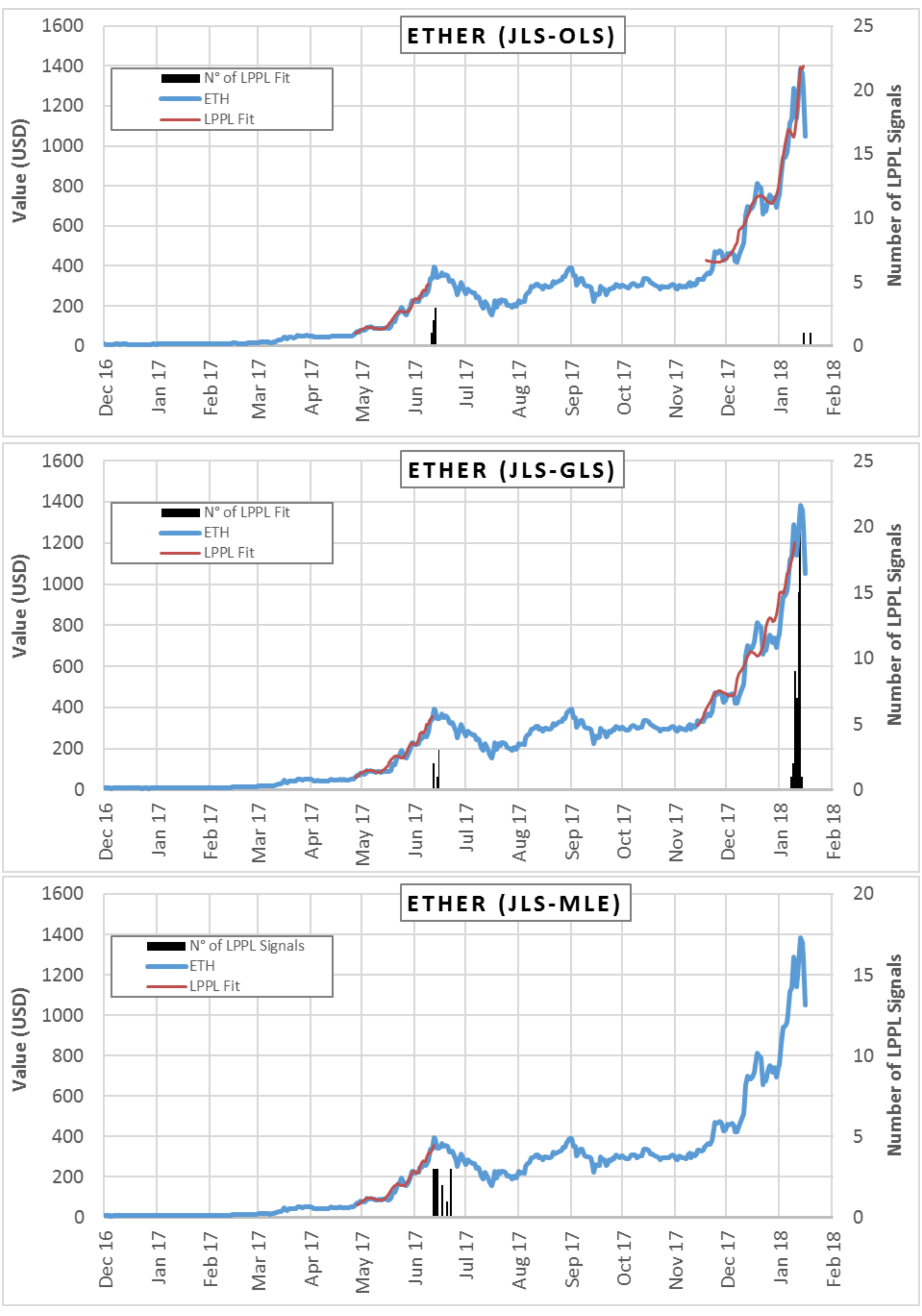

ETHER (JLS-OLS)
N° of LPPL Fit
ETH
LPPL Fit
Value (USD)
Number of LPPL Signals
1600
1400
1200
1000
800
600
400
200
0
25
20
15
10
5
0
Dec 16
Jan 17
Feb 17
Mar 17
Apr 17
May 17
Jun 17
Jul 17
Aug 17
Sep 17
Oct 17
Nov 17
Dec 17
Jan 18
Feb 18
ETHER (JLS-GLS)
N° of LPPL Fit
ETH
LPPL Fit
Value (USD)
Number of LPPL Signals
ETHER (JLS-MLE)
N° of LPPL Signals
ETH
LPPL Fit
Value (USD)
Number of LPPL Signals

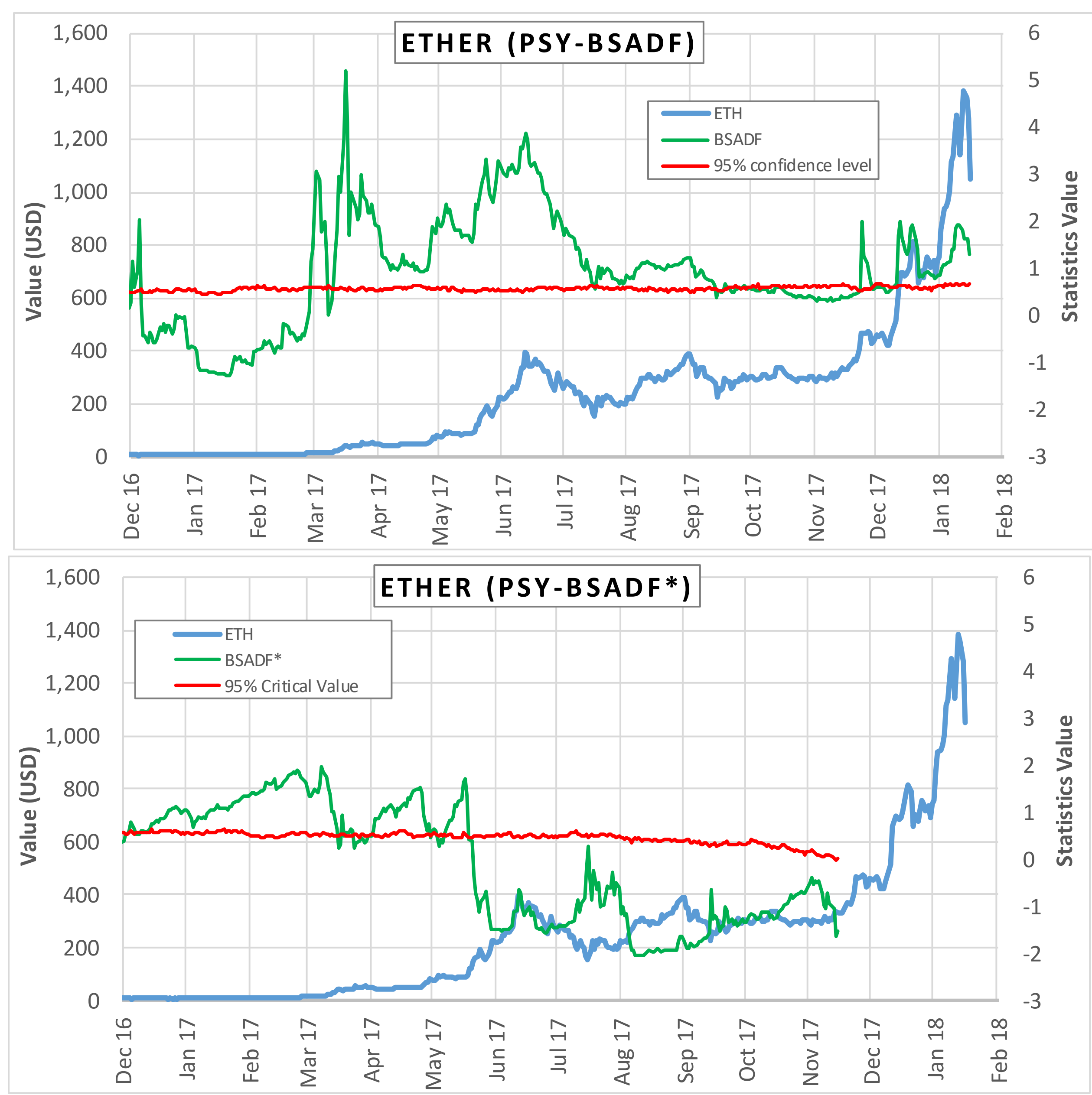

***Figure 3****: Ether results. Description as in Figure 2. Source: Etherscan.*

A second weaker signal is found by JLS-GLS only around 12th January 2018, anticipating the large crash observed in the same days, after the crazy rally of December 2017 (~200%). We notice that in the second half of December 2017 the ~30% Bitcoin crash caused a limited ~20% Ether slump lasting just a few days. In fact, the BTC/ETH correlation was nearly zero in that period (see Figure 1).

In Table 1 we present a detailed summary of cryptocurrency bubble signals found by each model.

| Time series | Bubble signals | | | | |
|---|---|---|---|---|---|
| | JLS | | | PSY | |
| | OLS | GLS | MLE | BSADF | BSADF* |
| Bitcoin | 15Dec2017 ± 9 (52)<br>19Jan2018 ± 7 (36) | 11Dec2017 ± 4 (29)<br>23Jan2018 ± 4 (15) | 14Dec2017 ± 3 (50) | 11Dec2016 - 06Jan2017 (**)<br>5May2017 - today (***) | 21Mar2017 - 2May2017 (**) |
| Ether | 12Jun2017 ± 2 (6) | 12Jun2017 ± 3 (6)<br>12Jan2018 ± 1 (56) | 15Jun2017 ± 4 (15) | 26Feb2017 - 20Jul2017 (***)<br>13Dec2017 - today (**) | 06Dec2017 - 14Mar2017 (**)<br>25Mar2017 - 23May2017 (*) |

***Table 1:*** *Summary of cryptocurrency bubble signals for each model applied. JLS bubbles are expressed as average signal date ± 1 standard deviation in days (number of signals); PSY bubbles are expressed as time window and confidence level: (*) = 50%-70% (low), (**) = 70%-90% (medium), (***) = 90%-100% (high).*

## 5. Conclusions

The growth of peer-to-peer exchanges and the blockchain technology has led to a proliferation of cryptocurrencies and to a massive increase in the number of investors who actually negotiate digital money. Cryptocurrencies trade at prices which is mainly driven by investor sentiment, becoming a potential source of financial instability.

We developed a quantitative bubble analysis methodology to detect financial bubbles and forecast their critical end time. In particular we developed in synergism two methods: the Johansen-Ledoit-Sornette (JLS) model, with three different calibrations (OLS, GLS, MLE) and robust global optimization methods, i.e. Genetic Algorithms, and the Philips-Shi-Yu (PSY) model, with two different statistical detection strategies (BSADF, BSADF*).

Our approach, applied to two of the most important cryptocurrencies, Bitcoin and Ether, consistently revealed strong bubble signals, often confirmed by actual crashes observed in the crypto-markets. In fact, Bitcoin shows typical hallmarks of a bubble phase in mid December 2017 and in the first half of January 2018, anticipating the large crashes observed thereafter. The Ether price dynamics reveals bubble evidence in mid June 2017, anticipating the crash observed on 12th June. A second weaker signal is found by JLS-GLS only around 12th January 2018, anticipating the crash observed in the same days, after the crazy rally of December 2017. These crashes can be related to recurring negative news of China and South Korea regarding restrictions for cryptos.

This paper provides insight into the high risk of speculative bubbles associated with cryptocurrencies, related to investor exuberance pumping market prices far away from their fundamental values, thus creating critical situations subject to possible crashes.

Our methodology is general and can be applied to virtually any financial time series, such as other cryptocurrencies, ICOs[2], forks[3], and related investing strategies characterising the present digital revolution.

[2] an innovative structure for raising funds to support new ideas and ventures based on digital tokens and blockchain-related technologies.
[3] forks are typically used to split existing blockchains to create new cryptocurrencies with different features (e.g. Bitcoin vs Bitcoin Cash).